 \documentclass[
 preprint,
 amsmath,amssymb,
 aps,
prb,
]{revtex4-2}

\usepackage{graphicx,bm,xcolor,microtype,multirow,amscd,amsmath,amssymb,amsfonts,physics,longtable,wrapfig,txfonts,soul}

\usepackage{adjustbox}

\usepackage[utf8]{inputenc}
\usepackage[T1]{fontenc}
\usepackage{txfonts}
\usepackage{siunitx}

\usepackage[
	colorlinks=true,
    citecolor=blue,
    breaklinks=true
	]{hyperref}
\begin{document}	
\preprint{APS/PR}

\title{
Stability and optoelectronic properties of oligothiophene  molecules confined  in boron-nitride nanotubes :  A many-body theoretical approach  }

\author{Xavier Blase}
\affiliation{ Univ. Grenoble Alpes, CNRS, Institut N\'{e}el, F-38042 Grenoble, France }
\email{xavier.blase@neel.cnrs.fr}
\author{Mauricio Rodriguez-Mayorga}
\affiliation{Laboratoire de Physique Th\'{e}orique, Université de Toulouse,
CNRS, Toulouse, France} 
  \author{Ivan Duchemin}
\affiliation{Univ. Grenoble Alpes, CEA, IRIG-MEM-L\_Sim, 38054  Grenoble, France}

\date{\today}

\begin{abstract}
We investigate the structural and optoelectronic properties of oligothiophene (nT) molecules encapsulated in boron-nitride (BN) nanotubes using density functional theory, many-body $GW$ and Bethe-Salpeter equation approaches.  We show that the binding energy is maximized for tube diameters of approximately 10~\AA, decreasing gradually for larger diameters. The nT molecules can slide along the tube with a corrugation potential smaller than room temperature thermal energy,   promoting their head-to-tail aggregation.
Regarding the electronic properties, we find that hybridization with the tube electronic states has a much smaller effect than that of screening, which can close the nT photoemission gap by as much as an eV. A simple model for the polarization of the BN tube demonstrates how these polarization effects decrease with increasing nT molecule length and BN tube diameter.
 Compared to the gas phase optical properties, structural relaxation, hybridization, and screening can redshift the absorption onset by up to 250 meV for isolated intercalated nTs. Additionally, the study of a head-to-tail sexithiophene dimer reveals an exciton-exciton interaction that splits the absorption onset into a lowest bright exciton, separated by approximately 75 meV from a dark peak. This suggests possible collective effects upon the formation of nT chains inside BN tubes. Our results confirm and clarify experimental data, mitigating the conclusion that insulating BN tubes just act as a protecting environment for encapsulated molecules.
\end{abstract}


\maketitle

\maketitle

\section{Introduction}

Encapsulation of organic dyes inside boron-nitride (BN) nanotubes has recently garnered significant experimental attention \cite{Niskanen2016,Allard2020,Datz2021,Badon2023,Jordan2023,AllardChemSocRev2024,Juergensen2025,Marceau2025,Marceau2026}, following earlier analogous studies involving carbon nanotubes  (for a recent Review, see Ref.~\citenum{AllardChemSocRev2024}). 
Encapsulating molecular systems inside nanotubes enables the formation of novel one-dimensional (1D) aggregates with specific properties. As compared to carbon nanotubes, BN analogs present the advantage of being wide gap insulators. As such, in most cases, the molecular states fall well within the BN tube's band gap,  exhibiting minimal  hybridization with the tube states. Besides templating novel molecular 1D arrangements, the BN tube walls may thereby protect the encapsulated molecule electronic properties. Stabilization of optical emission of  encapsulated dyes  against bleaching and blinking, spanning orders of magnitude in time scales, was demonstrated upon intercalation inside BN nanotubes  \cite{Allard2020}.  Among the typical molecular systems used for intercalation, the oligothiophene family has been the most studied at the experimental level  \cite{Allard2020,Badon2023,Juergensen2025,Marceau2025,Marceau2026}.

In this work, we investigate the structural and optoelectronic properties of oligothiophene (nT) molecules encapsulated in boron-nitride nanotubes (BNTs). We employ density functional theory (DFT) for structural properties and many-body $GW$ and Bethe-Salpeter (BSE) perturbation theories for electronic and optical properties.  Our results show that the binding energy of nT chains with BN tubes is maximized for radii of the order of $\sim$10~$\AA$, gradually decreasing for larger tubes. Additionally, we find that nT chains can slide along the tube with an energy barrier significantly lower than room temperature thermal energy. This confirms the possibility for nT molecules to form head-to-tail or side-by-side aggregates, depending on the tube radius. Although hybridization between nT molecular orbitals and BN tube electronic states is minimal, we demonstrate that long-range screening,  i.e. polarization effects,  can reduce the nT photoemission gap by up to 1~eV. Screening has a relatively smaller effect on optical absorption energies, still inducing a redshift of 100--150~meV. Combined with structural relaxation and hybridization effects during intercalation, this redshift can reach up to 240~meV for sexithiophene encapsulated in a 10~\AA-diameter BN(7,7) tube. Finally, we analyze the absorption properties of a head-to-tail sexithiophene dimer, evidencing the role of dipole-dipole interactions in the excited state.

\section{Stability of encapsulated \lowercase{n}T inside BN tubes }

Our ground-state DFT calculations for encapsulated nT@BN systems are performed with the {\sc{Siesta}} package \cite{Soler2002} allowing the study of large periodic BN tubes. 
We  employ  a double-zeta plus polarization basis (DZP) and the Vydrov and Van Voorhis (VV) nonlocal correlation functional \cite{Vydrov2010}  which accounts for  van der Waals (vdW) dispersive contributions. While several nonlocal functionals are available \cite{Dion2004,Lee2010,Klimes2009,Vydrov2010}, the VV one leads to similar characteristics for the nT@BN systems as compared to the Klime\v{s}, Bowler and Michaelides (KBM) functional  \cite{Klimes2009}, and the rather different B3LYP functional \cite{Becke1993,Lee1988} with  dispersion corrections \cite{Grimme2010} (see Supplemental Material~\cite{supplemental}). Given the absence of reference calculations for these specific systems, an extensive exploration of the \textit{pros and cons} of various vdW functionals stands beyond the scope of the present study. 
Using this computational set-up, the calculated BN bond length is 1.456~$\AA$,  which is marginally larger  than the 1.45~$\AA$ experimental value. 

We now examine the binding energy of a 4T molecule inside armchair and zigzag BNTs as a function of diameter. Our BN(n,n) armchair nanotube supercells include 12 BN(n,n) unit cells along the tube axis, totaling 48×n atoms and a length of $\sim$ 30.3 $\AA$, which is significantly larger than the maximum extent of the 4T molecule ($\sim$ 16.5~$\AA$). Alternatively, our BN(n,0) zigzag nanotubes include 7 BN(n,0) unit cells along the tube axis, amounting to 28×n atoms and a supercell length of 30.6~$\AA$.

\begin{figure}[t]
 \centering
\includegraphics[width=10.0cm]{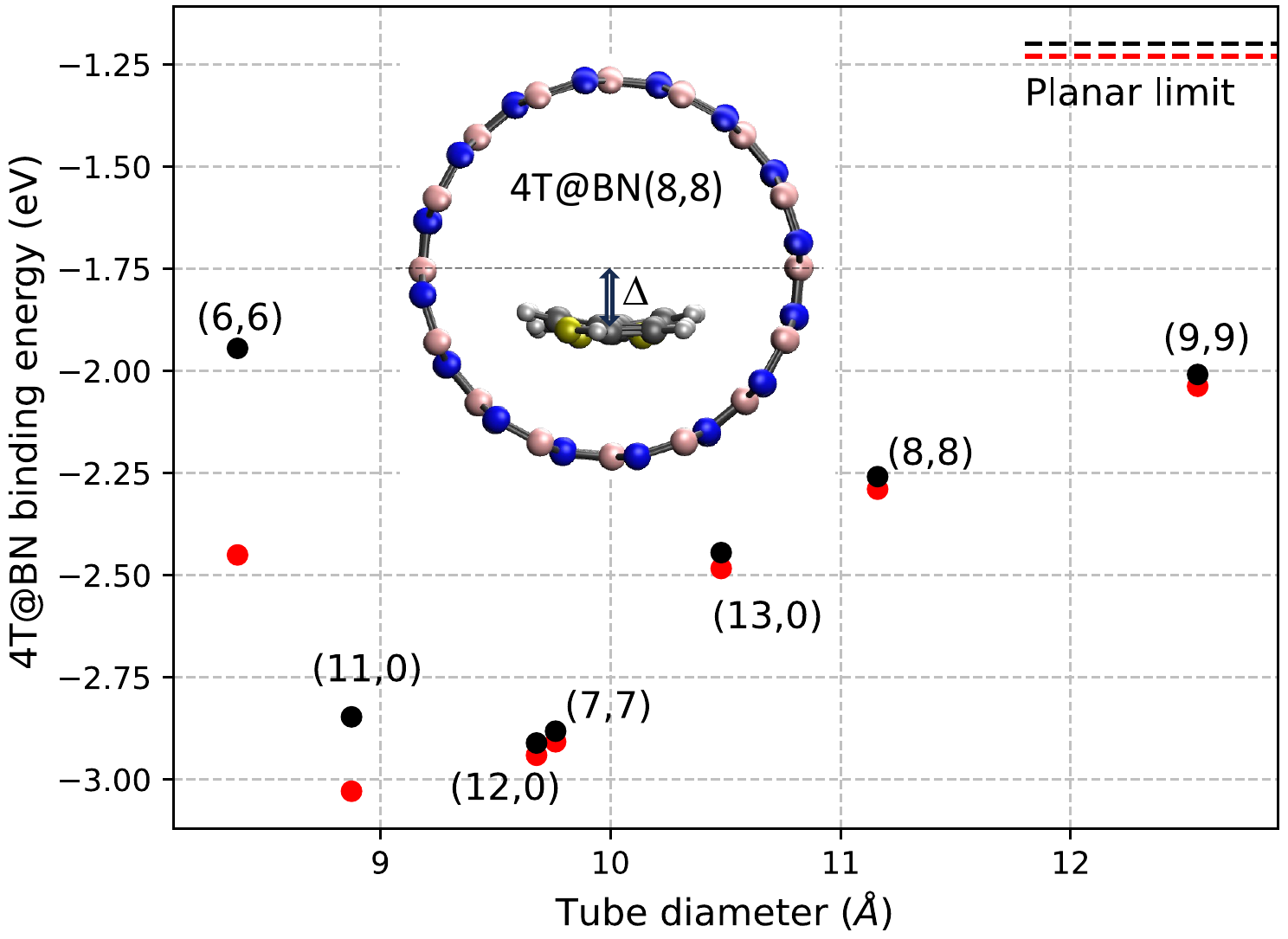}
	\caption{Binding energy (eV) for tetrathiophene (4T) in BN tubes with various diameters. The red dots indicate the binding energy using for the isolated 4T and BN tubes the frozen geometry of the interacting systems, while the black dots add the relaxation energy from these frozen geometries to the relaxed geometry of the isolated objects. Calculations with the VV functional at the DZP level. (Inset) Schematic representation of the 4T@BN(8,8) system. The 4T center-of-mass is shifted by   $\Delta\simeq 1.7~\AA$ with respect to the tube axis. }
	\label{fig:figstability}
\end{figure}

Our calculated binding energies are shown in Fig.~\ref{fig:figstability}, where two values are provided (red and black dots). From the energy of the relaxed interacting 4T@BN periodic systems, we subtract  the energy of the isolated 4T and BN subsystems in their \textit{frozen 4T@BN geometries}, which corresponds to the red dots. The  energies  of the isolated frozen subsystems  are calculated while accounting for ghost orbitals of the removed atoms to mitigate basis set superposition errors. We subsequently relax the isolated fragments to their ideal geometries. This additional relaxation energy yields the difference observed between the red and black dots. As expected, this difference is significant for the smaller tubes but becomes negligible ($\sim$30 meV) for tubes larger than $\sim$10~Å in diameter. The planar limit is derived by considering a 4T molecule deposited onto a periodic planar supercell, which corresponds to the ``{u}nfolded" BN(7,7) tube supercell considered earlier (336 B/N atoms). The molecule-to-planar \textit{h}-BN distance amounts to $\sim$3.30~$\AA$. 

Our results indicate a maximal binding energy for a diameter slightly smaller than 10~$\AA$, specifically for BN(7,7) or BN(12,0) nanotubes in our calculations. As expected, the binding energy decreases rapidly for smaller diameters and more gradually for larger diameters.  High-resolution transmission electron microscopy (HRTEM) experiments indicate that single 6T molecules can be encapsulated  in BNTs with diameters as small as 0.9~nm \cite{Allard2020}.  In another study \cite{Badon2023}, the distribution of BN nanotube diameters incorporating 6T molecules was found to peak between 10~$\AA$ and 20~$\AA$. This is very consistent with our calculated data.
Our results for the 6T encapsulated inside a BN(7,7) tube yield a binding energy of about 4.43~eV, instead of  2.91~eV for the encapsulated 4T@BN77. This suggests that the binding energy scales approximately linearly with the number of thiophene monomers. As such, our findings for the 4T molecule can be extrapolated to the 6T molecule.

Our results are further supported by recent semi-empirical Lennard-Jones potential calculations, which predict a maximum binding energy for single nT molecules in BNTs with diameters of approximately 9.4~$\AA$ \cite{Elouardi2025}. However, the semi-empirical calculations predict significantly larger binding energies ($\sim$8.4~eV for 6T)  as compared to our \textit{ab initio} results (4.4~eV).

\section{Easy sliding of oligothiophenes inside BN tubes}

We select the   BN(7,7) nanotube using periodic boundary conditions with the supercell described above, and consider again the case of 4T molecules. We pull the molecule rigidly along the tube axis by steps of 0.2 $\AA$ over one BN tube unit cell, relaxing the full system  at each step. The energy does not change by more than a few meV when sliding the 4T molecule, indicating a very small corrugation potential. This is much smaller than room temperature, assigning 3$k_B$T/2 thermal energy to each  atom of the molecule. As such, the thiophene molecules are expected to slide easily along the tube axis. This confirms the experimental observation that head-to-tail chains of 6T form inside BN tubes not large enough to accomodate side-to-side nTs  \cite{Allard2020,Badon2023,Juergensen2025}.

\begin{figure}[t]
\includegraphics[width=12cm]{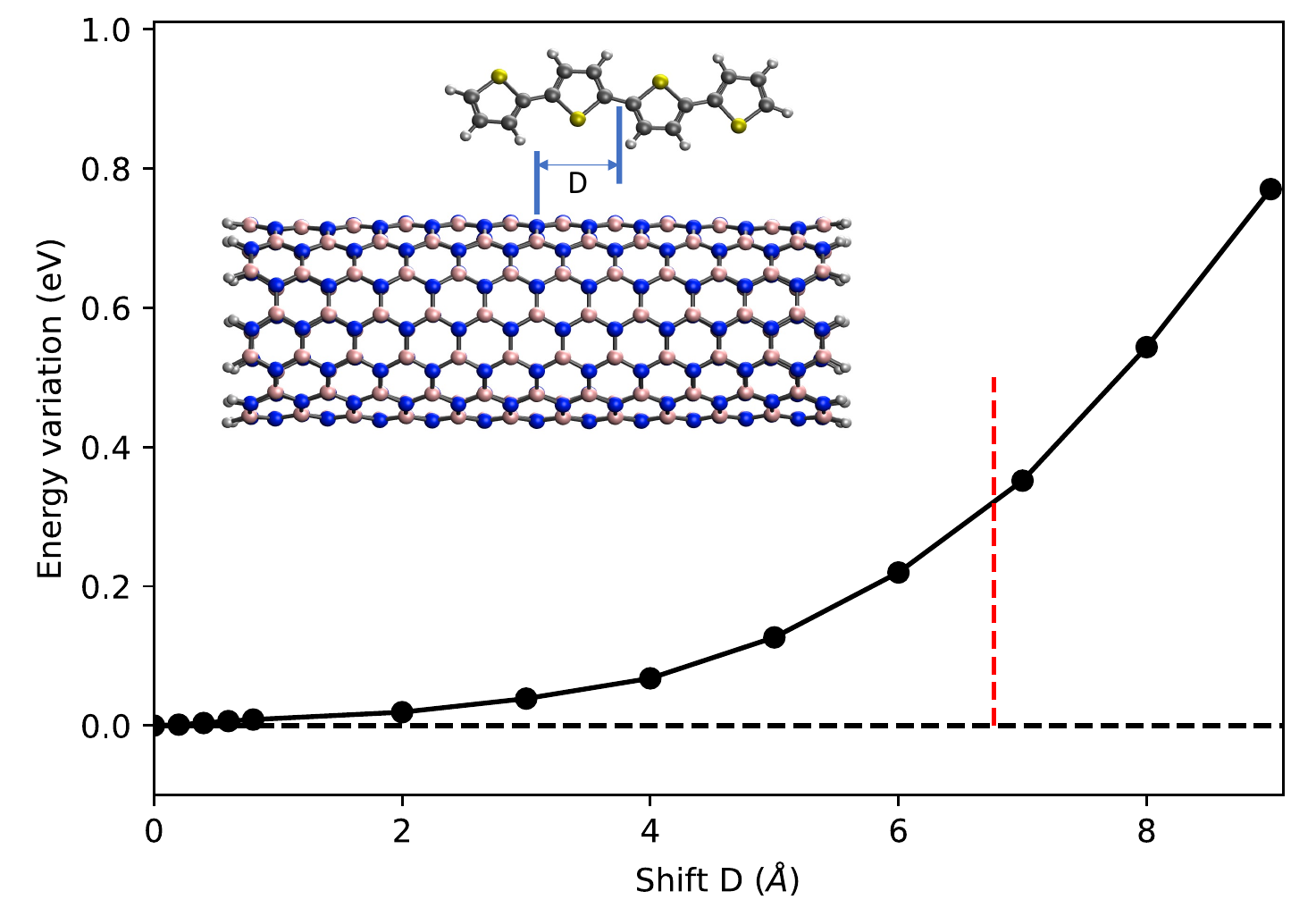}
	\caption{ Evolution of the energy of the 4T inside a finite size BN(7,7) tube as a function of the pulling distance D. The distance D represents the pulling distance along the tube axis, with for D=0~$\AA$ the 4T and tube centers being aligned. The vertical dashed red line indicates the pulling distance D at which the right-most 4T carbon atom  coincide with the tube right-most B or N atoms. (Inset) Schematic representation of the 4T sliding inside the finite size BN(7,7) tube. The 4T is placed outside the tube for sake of visibility.   }
	\label{fig:figsliding}
\end{figure}

As another interesting exercise, we consider an open-ended BN(7,7) tube with edges passivated by hydrogen. We pull rigidly the 4T molecule along the tube axis (see Inset Fig.~\ref{fig:figsliding}). While the energy hardly changes in the vicinity of the D=0 pulling distance, the binding energy reduces as expected when the 4T molecule starts leaving the tube. The rather steep increase of the energy when the molecule leaves the tube measures the driving force for the \textit{barrier-less} incorporation of oligothiophene in open ended tubes. How such a driving force will be affected by the presence of a solvent is a question that we do not address. 

\section{ Photoemission and optical gaps }

We now study the photoemission and optical properties of encapsulated nTs using the many-body $GW$ \cite{Hedin1965,Strinati1982,Hybertsen1986,Godby1988,ReiningBook,Golze2019}  and Bethe-Salpeter equation (BSE) \cite{Salpeter1951,Csanak1971,Strinati1982b,Strinati1988,Benedict1998,Rohlfing1998,Albrecht1998,Blase2018,Blase2020} formalisms. The many-body $GW$ approach, where $G$ stands for the one-body Green's function and $W$ is the screened Coulomb potential, provides electronic energy levels in much better agreement with photoemission experiments as compared to standard Kohn-Sham DFT. We refer the reader to extensive benchmarks concerning $GW$ calculations on organic systems  \cite{Setten2015,Rangel2016,Kaplan2016,Knight2016}. In particular, the electronic relaxation energies associated with the addition/removal of an electron in a photoemission experiment, including the response from a dielectric environment, is fully accounted for within linear response (see e.g. Refs.~\citenum{Li2018,Baumeier2014}). 
Similarly, numerous benchmark calculations have demonstrated the ability associated with the Bethe-Salpeter formalism to deal accurately with optical excitations of various nature, from extended Wannier to localized Frenkel or charge-transfer excitations  \cite{Baumeier2012,Bruneval2015,Jacquemin2015,Gui2018,Jacquemin2017,Knysh2024}. As for the $GW$ formalism, the dielectric response of the environment to an optical  excitation on a central subsystem of interest, and the related shifts in energy, can be fully accounted for  \cite{Baumeier2014,Duchemin2018}.

In what follows, the encapsulating \textit{h}-BN nanotube is considered as the ``environment", and the central subsystem of interest is the encapsulated nT. Our main goal in this Section is to understand the impact of the nanotube electronic degrees of freedom, via hybridization and screening, onto the photoemission gap  and onset of optical absorption associated with the encapsulated molecule. Even though a large gap insulator, we show that the BN nanotube induces a non negligible renormalization of the optoelectronic properties of the encapsulated nTs, mitigating the common understanding that the BN tube only acts as a protection for the intercalated molecule. This renormalization is mainly due to screening, with weak hybridization, so that the highest occupied (HOMO) and lowest unoccupied (LUMO) molecular orbitals, together with the associated lowest bright excitation,  remain strongly localized onto the intercalated molecule in the coupled nT@BN systems.   

Our many-body calculations are performed with the {\sc{beDeft}} (beyond-DFT) code \cite{Duchemin2020,Duchemin2021} developed for molecular systems in the gas phase or ``{e}mbedded" in large scale dielectric environments treated at a fully \textit{ab initio} level  \cite{Amblard2023,Amblard2024,amblardphd}. Since our many-body code only tackles finite-size systems using Gaussian basis sets, we start with DFT calculations performed with the {\sc{Orca}} code  \cite{Neese2009,Neese2022}. We favor for structural relaxations the B3LYP functional
\cite{Becke1993,Lee1988} with D3 dispersion corrections  \cite{Grimme2010}, together with the efficient 6-31G(d) basis set and the associated geometric counterpoise corrections as benchmarked in Ref.~\citenum{Kruse2012}.
We emphasize that we aim to relax  systems containing up to 380 atoms. With such a set-up, we find the BN bond length to amount to 1.447~$\AA$, in close agreement with the 1.45 $\AA$ experimental value. As shown in the  SM  \cite{supplemental}, the geometry of the 3T@BN(7,7) relaxed system that we study here below is in excellent agreement with that obtained with the {\sc{Siesta}} code and the VV   functional  used in the first Section of this study. 

Concerning our many-body calculations of the electronic and optical properties, following numerous benchmarks on molecular systems,  the needed input Kohn-Sham  eigenstates are generated with the PBE0 functional \cite{Perdew1996,Ernzerhof1999,Adamo1999} containing an increased amount of exact exchange. We further adopt for the nT molecules the larger def2-TZVP basis set \cite{Weigend2005} with its associated def2-TZVP-RIFIT auxiliary basis  \cite{Weigend1998}. The BN tube is described at the def2-SVPD level  \cite{Rappoport2010}, a basis set specifically designed for response properties. This basis was chosen having in mind that we are mainly interested in the effect of the tube polarization (screening) on the nT electronic excitation, and not in the optoelectronic properties of \textit{h}-BN systems that are already well documented at the many-body level   \cite{Blase1995,Arnaud2006,Wirtz2006,Attaccalite2011,Attaccalite2018,Cannuccia2019}.   The related DFT Kohn-Sham eigenstates serve as a starting point for the many-body calculations. Our $GW$ calculations are performed with the eigenvalue-self-consistent (ev$GW$) scheme  \cite{Shishkin2007,Blase2011,Kaplan2016,Rangel2016}. At the BSE level, we go  beyond the Tamm-Dancoff approximation, including all occupied and unoccupied states.  As a first validation, the obtained ground-state B3LYP geometries and related BSE/ev$GW$@PBE0 calculations on 3T and 6T gas phase molecules lead to  lowest singlet (S$_1$)  absorption energies of 3.42 eV and 2.67 eV. Our absorption energy for the smallest 3T is in good agreement with the available 3.57 eV coupled-cluster (CC2) calculation  \cite{Fabiano2005,fourthio}. Further, our calculated energies are in good to reasonable agreement with the  3.49 eV and 2.85  eV experimental  data. \cite{Lap1997}  Beyond absolute values, we are mainly interested in the evolution from the gas phase as a consequence of intercalation in BN tubes. The nTs lowest $S_1$ optical excitation is bright with a transition dipole that is non-zero only for an electric field aligned with the molecule principal axis. We refer the reader to Ref.~\citenum{Cocchi2015} for a study at the BSE level of the evolution of the nTs optical spectra from the gas phase to the crystalline form. 

We start by exploring the properties of the 3T and 6T molecules intercalated in a BN(7,7) tube. Our results are compiled in Table~\ref{tab:table1}. We  focus  first on structural relaxation effects. Oligothiophenes are known to easily distort, in relation in particular with very-low energy twisting modes between thiophene monomers. Such twisting modes disturb the conjugation along the chain, leading to an opening of the band gap  \cite{Zade2007}. This is the first effect we observe. While non-planar in the gas phase, the nT molecules become planarized when deposited on a flat  \textit{h}-BN substrate due to van der Waals (vdW) interactions. Constrained relaxation forcing the nT atoms to remain strictly in the same plane leads to very similar effects. This planarization leads to a 0.13 eV/0.09 eV closing of the PBE0 band gap for the 3T/6T. We emphasize that we consider here the isolated nT in their planarized geometries, namely this does not include the effect of hybridization with the BN orbitals. This \textit{gap closing due to geometrical effects} propagates to the ev$GW$ photoemission gap and BSE lowest optical absorption energy. Planarization leads in particular to a 0.12/0.09 eV lowering of the 3T/6T lowest singlet S$_1$ energy. 

 Due to the tube curvature, the geometry of nT molecules incorporated in the BN(7,7) small diameter tube is more complex, with  local minima that depend on the initial nT configuration used for relaxation. Starting with the non-planar 3T gas phase configurations, the twists do  not disappear and the thiophene monomers seem to "stick" to the closest BN wall portion. Starting from a planarized 3T, the intercalated 3T  remains close to planar with a total energy within $\sim$50 meV from the twisted intercalated conformer, an energy difference lower than room-temperature energy. Having in mind that for larger tubes the planar structure is expected to prevail, following the planar \textit{h}-BN limit,  we adopt the planar isomers for the 3T@BN77 and 6T@BN77 systems.

\begin{table}[t]
\begin{center} 
\begin{tabular}{ c|c|c|c } 
                             &  PBE0 gap & ev$GW$ gap & BSE S$_1$  \\ \hline  
3T gas                       &  3.84    & 6.89       &  3.42       \\  
3T gas (@BN planar geom)     &  3.71    & 6.72       &  3.30  \\      
3T gas (planar constrained)  &  3.71    & 6.72       &  3.30  \\ 
3T gas (@BN77  geom)         &  3.76  & 6.78       &  3.34      \\ 
3T@BN77(10~\AA)              &  3.71  &  6.08      &  3.15  \\   
3T@BN77($\infty$)            &  3.71  &  5.78      &  3.15  \\ \hline 
6T gas                       &  2.95  &  5.47      &  2.67    \\
6T gas (@BN planar geom)     &  2.87  &  5.36      &  2.59  \\
6T gas (planar constrained)  &  2.86  &  5.34      &  2.58     \\
6T gas (@BN77  geom)         &  2.88  &  5.37      &  2.60            \\
6T@BN77($\infty$)            &  2.85  & [4.60]    &    [2.43]      \\
    \end{tabular}
\end{center}
\caption{ Calculated 3T/6T    PBE0  and ev$GW$@PBE0 gaps and BSE lowest singlet (S$_1$) absorption energy (eV) as a function of geometry and environment.  The ``gas" data indicate non-interacting nTs obtained for various geometries.  The nT@BN77($\infty$) extrapolate screening effects to infinite size  tubes (see Fig.~\ref{fig:ximodel}). Numbers in bracket indicate that the gap closing by hybridization at the Kohn-Sham level has been accounted for perturbatively  (see text).  } 
\label{tab:table1}
\end{table}

We now look at the optoelectronic properties of the intercalated nT@BN systems, starting with a small 3T@BN77(10~$\AA$) system, namely a 3T molecule intercalated in a 10~$\AA$ long BN(7,7) tube section. Tube edges are passivated by hydrogens. This is a minimal tube in the sense that the 3T length is of 10.3~$\AA$ from outmost carbon-to-carbon atoms, a length equivalent to the 10~$\AA$ separation between the outermost BN rings of the tube section. As discussed below, this tube section is not long enough to converge the photoemission ev$GW$ gap of the intercalated nT, but the nT optical properties are already fully converged. 

The analysis of the data in Table~\ref{tab:table1} indicates that the interaction with the BN tube at the DFT Kohn-Sham level (electrostatic and hybridization) closes the 3T Kohn-Sham gap by $\sim$50~meV, from 3.76 eV to 3.71 eV. We compare here   the   3T gas (@BN77 geom) system, namely the 3T in the gas phase but frozen in its intercalated geometry,  to the fully interacting  3T@BN77(10 Å)  complex.   The hybridization with the tube orbitals at the Kohn-Sham level cancels here the effect of relaxation away from the perfect planar geometry. Such a full cancellation is certainly accidental. 

The evolution of the ev$GW$ gap from the  3T gas (@BN77 geom) system to the interacting  3T@BN77(10 $\AA$) is much larger, amounting to 700 meV, from 6.78 eV to 6.08 eV. This is the effect of nonlocal screening, namely the dielectric response of the tube to an electronic excitation on the 3T molecule. This is fully accounted for at the $GW$ level. The onset of optical absorption  (S$_1$ energy) is also redshifted, but  by a reduced 190 meVs, from 3.34 eV to 3.15 eV.  This reduced impact of the tube dielectric response on the 3T optical properties can be related to the fact that an optical excitation is associated with a neutral charge reorganisation, while photoemission fully charges the nT subsystem. As compared to the fully relaxed 3T in the gas phase, the onset of optical absorption reduces by 0.27 eV, from 3.42 eV to 3.15 eV, as a combined effect of structural relaxation, hybridization and screening.

We now address the case of longer BN tube sections.  To deal with significantly larger systems, and since hybridization at the Kohn-Sham level is small, we adopt a ``{f}ragment" approximation  \cite{Fujita2018,Liu2019,Fujita2021,Tolle2021,Amblard2022,Amblard2023}. Namely, we decouple the nT and tube electronic degrees of freedom, leading to a block diagonal independent-electron  susceptibility. Within the fragment approximation, the 3T@BN77(10 $\AA$) system ev$GW$ gap and BSE S$_1$ energy amounts to 6.14 eV and 3.22 eV, instead of 6.08 eV and 3.15 eV allowing full hybridization between the molecule and the tube. Most of the difference can be explained already by the 50~meV closing of the gap by hybridization at the Kohn-Sham level. The small remaining 10/20 meV can be considered as a measure of the impact of the fragmentation (block diagonalization) of the independent-electron susceptibility operator, together possibly with the neglect of mixing between optical excitations on the molecule and on the BN tube at the BSE level.

To further allow studying larger tube sections, the tube susceptibility is calculated using a real-space imaginary-time approach \cite{Duchemin2021} with a plasmon-pole approximation \cite{Duchemin2024} for the dynamical nature of screening. Following a recent study  \cite{Amblard2022,Amblard2023}, we  also build long BN tubes susceptibility operators by compressing and translating the corresponding operator associated with  fragments 20 $\AA$  or 28 $\AA$-long  \cite{hydrogens}.  These approximations are shown to be accurate at the few meV level in the SM~\cite{supplemental} and we focus here below on the results. 

\begin{table}[t]
\begin{center} 
\begin{tabular}{ c|cccc} 
                      & \;  Gap \; &   \;  S$_1$  \;       & \; $\langle H_0 \rangle$ \; & \; -$\langle  W \rangle$ \;   \\ \hline  
        3T gas (@BN77  geom)           & 6.78   &  3.34  & 7.19 & -4.68   \\ 
        3T@BN77(10~\AA)                & 6.14   & 3.22      & 6.45     & -3.98   \\
          3T@BN77(20~\AA)                &  5.97 &  3.22     & 6.27    &  -3.79   \\
          3T@BN77(28~\AA)                &  5.90 & 3.22     & 3.22   & -3.73    \\
         3T@BN77(20$\times$3~\AA)       & 5.85   & 3.22     & 6.15    & -3.67  \\
        3T@BN77(28$\times$3~\AA)        & 5.844  & 3.22     & 6.148 & -3.659 \\
        3T@BN77(20$\times$5~\AA)       & 5.842  & 3.22     & 6.146   & -3.657  \\ \hline
        6T gas (@BN77  geom)           &  5.37     &  2.60      & 5.75 & -3.79   \\
        6T@BN77(28$\times$3~\AA)       & 4.63      &  2.46    & 4.89   &  -2.95    \\
        \end{tabular} 
\end{center}
\caption{   $GW$ HOMO-LUMO gap and BSE $S_1$ energy (eV) for the   3T    in   BN(7,7) tube sections with increasing lengths. Calculations are here performed within the ``fragment" approximation, neglecting nT/tube wavefunctions hybridization, focusing on long-range screening. The notation (e.g.) $20 \times 5~\AA$ indicates that the susceptibility matrix of a 100~$\AA$ long BN tube is constructed by blocks by replicating 5 times the full susceptibility of a $20~\AA$ tube section.  We keep data at the meV level for sake of comparisons when needed. The $\langle H_0 \rangle$ and -$\langle  W \rangle$ columns indicate the non-interacting and electron-hole interaction contributions (see text) to the   $S_1$ energy. } 
\label{tab:table2}
\end{table}

\begin{figure}[t]
\includegraphics[width=12cm]{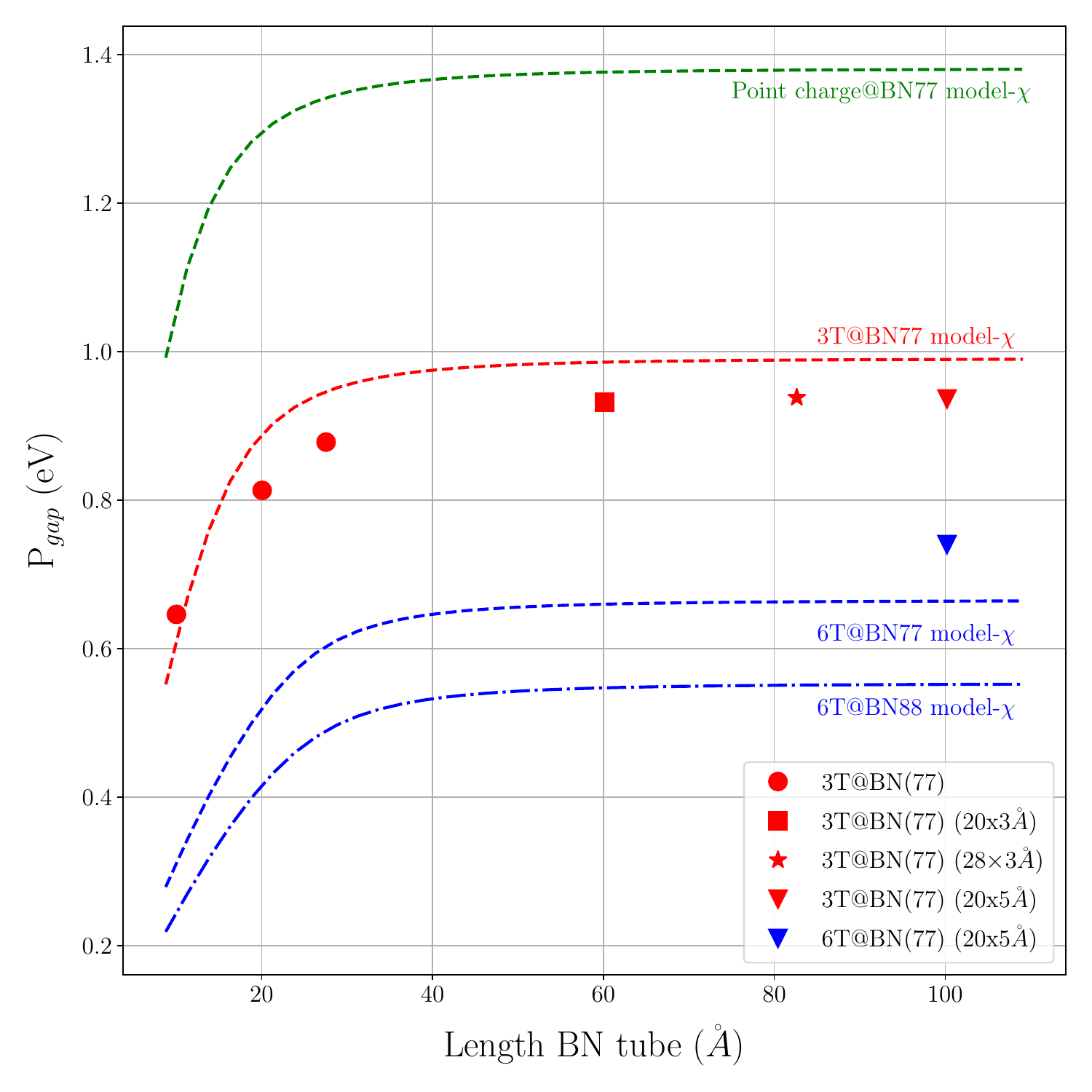}
	\caption{ Evolution with BN tube length of the gap polarization energy for  encapsulated 3T and 6T  molecules. This energy measures the reduction of the nT gap due to screening by the BN tube. The red/blue  symbols (circles, square, star, triangle) are the 3T/6T@BN77(L) ev$GW$ gap data from Table~\ref{tab:table2}  to which is subtracted the 3T/6T gas (@BN77  geom) ev$GW$ gap. The full lines are the output of the susceptibility model for \textit{h}-BN in the case of a centered unity point charge (green dashed), 3T (red dashed) and 6T (blue dashed) in   BN(7,7).  The dot-dashed blue line are the results of the model for the 6T@BN(8,8) system.  }
	\label{fig:ximodel}
\end{figure}

The evolution of the 3T@BN77(L)  ev$GW$ gap and BSE S$_1$ energies as a function of   tube length (L) is reported in Table~\ref{tab:table2},  indicating that the 3T@BN(7,7) ev$GW$ photoemission gap converges with tube length for L$\sim$100 $\AA$. We observe that replicating by blocks the susceptibility of a 20 $\AA$ or 28 $\AA$ tube section, to build the susceptibility of long tubes, leads to the same converged values at the very few meV level. This confirms the accuracy of constructing the non-interacting susceptibility of very long BN tubes by simply ``patching" together non-interacting susceptibility blocks corresponding to shorter tube sections. We further plot in Fig.~\ref{fig:ximodel} with red symbols the corresponding polarization energy that we define as the difference between the 3T@BN77(L) gap and that of the gas phase 3T molecule frozen in its intercalated geometry. This polarization energy measures the effect of the tube polarization in response to adding an electron or a hole on the 3T molecule in a photoemission experiment. This electronic reorganization energy stabilizes the added hole (smaller ionization potential) and electron (larger electronic affinity) by roughly the same amount, closing thus the photoemission gap. 

The \textit{ab initio} data are compared to a simple model (red dashed line, Fig.~\ref{fig:ximodel}) where all B/N atoms are associated with an onsite anisotropic susceptibility model adjusted to reproduce the polarizability of \textit{h}-BN flakes as calculated in Ref.~\citenum{Amblard2022}. The HOMO and LUMO charge distributions, that generates the polarization response, are taken to be the Mulliken charges as generated by the Orca code. The model is described in the Appendix below. Even though rudimentary, the model follows nicely, at least qualitatively, the calculated \textit{ab initio} data, confirming the rate of convergence with length of the polarization energy associated with the photoemission gap. 

In contrast, the analysis of the S$_1$ onset of absorption reveals that   convergence with tube length is achieved for the ``{s}mallest" (191 atoms) 3T@BN77(10 $\AA$) system. As explained above, an optical excitation is only associated with a neutral charge reorganisation, inducing a weaker reaction from the tube electronic degrees of freedom. It is interesting to understand from a different standpoint why the encapsulated 3T ev$GW$ photoemission gap may keep decreasing with BN tube length, while its optical gap does not change  beyond L$\simeq$10 $\AA$. We provide in Table~\ref{tab:table2} the evolution of the optical excitation energy without electron-hole interaction (the $\langle H_0  \rangle$ contribution) together with the strength of the attractive screened electron-hole interaction (the  $\langle-W \rangle$ contribution). As expected, the $\langle H_0  \rangle$ energy decreases following the photoemission gap, \cite{h0energy} but the electron-hole interaction reduces as much with increased screening. Both effects compensate exactly. 

We finally analyze the data associated with the larger 6T molecule encapsulated in the BN(7,7) system. In the fragment approximation  (Table ~\ref{tab:table2}), taking the 6T@BN77(28$\times$3 $\AA$) system as the converged reference, the effect of screening reduces the ev$GW$ gap from 5.37 eV to 4.63 eV, and the BSE S$_1$ energy from 2.60 eV to 2.46 eV. The reduction of the photoemission gap due to screening by the BN tube    amounts thus to 0.74~eV at the ev$GW$ level (see blue symbol and dashed blue model in Fig.~\ref{fig:ximodel}). This is somehow smaller than what was obtained for the 3T. A simple image charge model allows  understanding this reduction with nT length. \cite{imagecharge} On the contrary, the response to a point charge is larger (see green dashed line, Fig.~\ref{fig:ximodel}). 

The Kohn-Sham DFT calculation for the 6T@BN(28 $\AA$) system reveals a 30~meV closing of the gap by hybridization. This is again significantly smaller than screening effects. Adding these hybridization-related 30~meV to screening effects, the  final ev$GW$ gap and S$_1$ energies amount to  4.60 eV and 2.43 eV, respectively (numbers in brackets in Table~\ref{tab:table1}). As compared to the values obtained for the 6T relaxed in the gas phase, this represents a reduction of the photoemission  gap by 0.87 eV, from which 100 meV comes from structural relaxation, and 30 meV from hybridization in the ground state. As for the optical absorption onset, the redshift amounts to 240 meV, with 70 meV originating from structural relaxation and 30 meV from hybridization. 

We conclude this exploration by considering the case of the 6T@BN(8,8) system using the tube susceptibility model. The comparison of the blue dashed and blue dot-dashed lines in Fig.~\ref{fig:ximodel}  indicates that screening effects are reduced with increased tube diameter. Even though the number of B/N polarizable atoms increases linearly with tube diameter for a given tube length, the source-charge interaction with induced-dipoles on B/N atoms decays much faster with tube diameter  \cite{scaling}. As such, the magnitude of the screening effects obtained for the BN(7,7) tube should be considered as an upper-bound. 

\section{ Optical properties of a  head-to-tail dimer }

\begin{figure}[t]
\includegraphics[width=8.6cm]{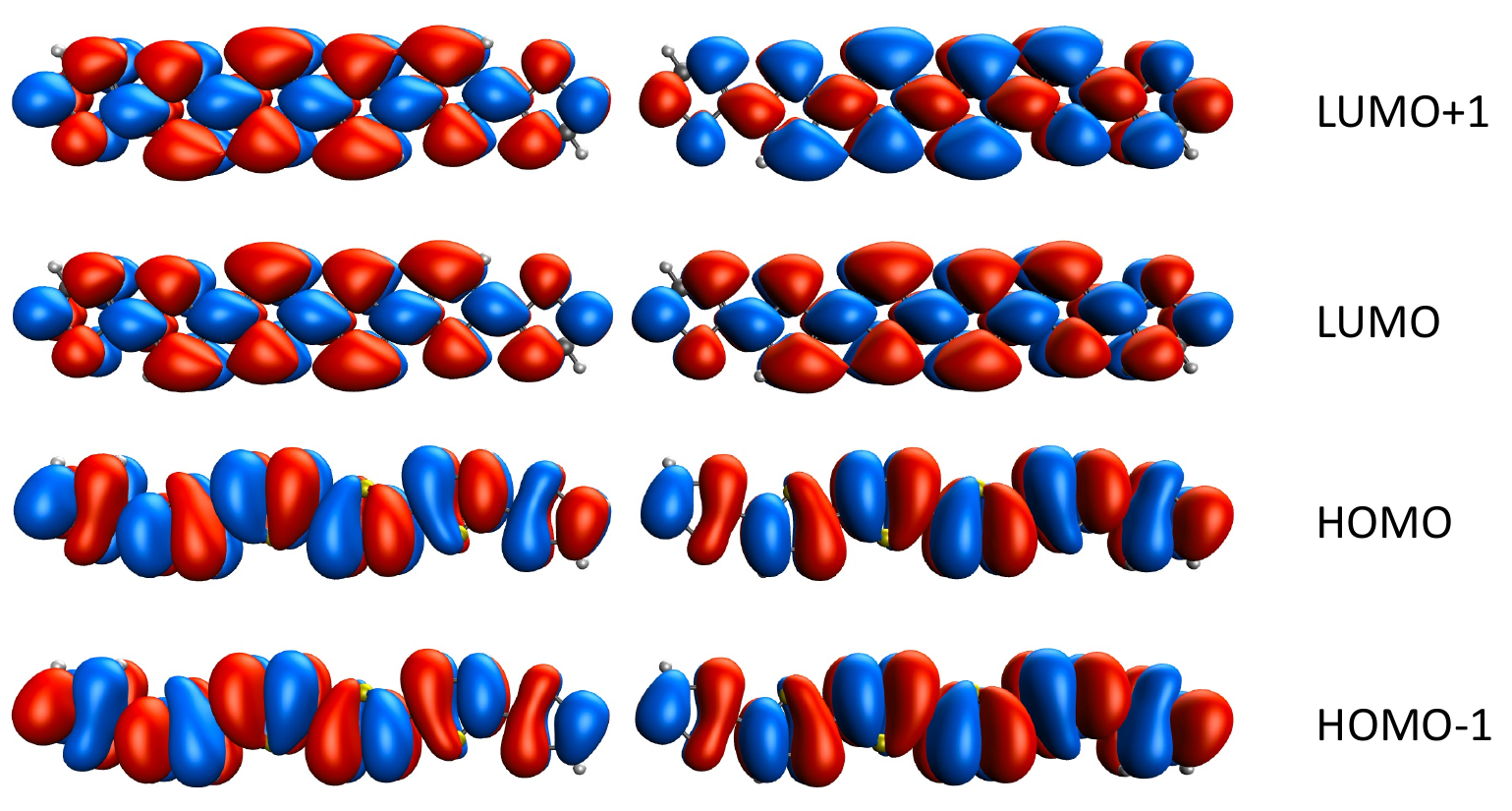}
	\caption{ 6T dimer frontier molecular orbitals ($T_z =26.0 ~\AA$).  }
	\label{fig:dimerwfns}
\end{figure}

We close this study  by considering a model head-to-tail ``{J}-aggregate" sexithiophene dimer. Our goal is to explore the possibility that  collective effects in the optical excited state associated with the formation of 6T chains, may significantly affect the onset of absorption as observed experimentally  \cite{Juergensen2025}. While the lattice vector of encapsulated one-dimensional 6T chains was found to be experimentally 25.10~$\AA$, we first construct a model dimer by translating rigidly a planar 6T molecule by  $T$=26 $\AA$ along the monomer  backbone   direction (see Fig.~\ref{fig:dimerwfns}). For this distance, the splitting of the HOMO and LUMO levels amounts to a vanishing $\sim$3 meV, allowing to focus specifically on the exciton-exciton interaction. The dimer frontier MOs are mainly constructed as linear combinations of the monomers frontier MOs (see Fig. \ref{fig:dimerwfns}): 
\begin{align}
    | L+1 \rangle &\simeq (  | L_1 \rangle - | L_2 \rangle ) / \sqrt{2} \nonumber \\
    | L \rangle &\simeq ( | L_1 \rangle + | L_2 \rangle ) / \sqrt{2} \nonumber \\
    | H    \rangle &\simeq ( | H_1 \rangle + | H_2 \rangle ) / \sqrt{2}  \nonumber \\
    | H -1  \rangle & \simeq ( -| H_1 \rangle + | H_2 \rangle ) / \sqrt{2} \nonumber  
\end{align}
where e.g. $|H\rangle$ and $| L \rangle$ are the HOMO and LUMO of the dimer, while $| H_1 \rangle$ and $| L_1 \rangle$ are the HOMO and LUMO of the "first" 6T composing the dimer. Here $|H_1 \rangle$ and $|H_2 \rangle$ have the same phase, as well as $|L_1 \rangle$ and $|L_2 \rangle$. 

The BSE calculations yield two lowest excitonic eigenstates \textit{split by 75~meV}. As such, even in the limit of negligible overlap between the monomers dimer, the exciton-exciton interactions can be quite sizable. The lowest is bright while the second is dark, an ordering specific of head-to-tail J-aggregates  \cite{Kasha1963,Hestand2018}.  The analysis of the BSE  excitations shows that the dimer  excitonic eigenstates read approximatively:  
\begin{align}
  | S_1^{\text{bright}}  \rangle & \simeq \frac{ | H \times L \rangle - | (H-1) \times (L+1) \rangle }{ \sqrt{2}  }  \nonumber \\
  | S_1^{\text{dark}}  \rangle &\simeq  \frac{ |( H-1) \times L \rangle - |  H  \times (L+1) \rangle }{ \sqrt{2}  }  \nonumber
\end{align}
The transition dipole along the z-direction associated with the lowest bright excitation is thus equal to: $ \langle H_1 L_1 + H_2 L_2| {\hat z}\rangle /\sqrt{2}$, with the 6T dimer elongated in the (z) direction.  As such, the dimer transition dipole moment is enhanced by $\sqrt{2}$ as compared to the monomer, and the associated oscillator strength by a factor 2.  This is consistent with the \textit{ab initio} results showing that
the calculated BSE oscillator strength for the bright S$_1$ amounts to (f=4.9) to be compared to (f=2.3) for the isolated monomer S$_1$. 
The transition dipole for the dark excitation is zero since proportional to $\langle H_1 L_1 - H_2 L_2| {\hat z}\rangle  $. 
As compared to the monomer lowest singlet excitation, the dimer bright S$_1$ is redshifted by 43~meV.   

Repeating the same calculation but incorporating the 6T dimer inside a BN(7,7) tube in the infinite limit, we find that the splitting between the bright and dark states reduces to about 50 meV. This is consistent with the idea that the tube screens the dipole-dipole interaction. For tubes larger than the small BN(7,7) considered here, one may thus expect a splitting standing in between 50 meV and 75 meV.  Even though preliminary, these results do not stand in favor of an exaltation of excitonic interactions along 6T chains, as suggested in Ref.~\citenum{Juergensen2025},  thanks to the tube electronic degrees of freedom. We emphasize however that in our embedding scheme, the direct coupling between nT and BN tube excitations at the BSE level is not allowed.  

    
Considering now the experimental 25.10 $\AA$ lattice parameter, the HOMO and LUMO splittings amount to 21 meV and 26 meV, respectively, at the $GW$ level. The resulting splitting between the (lowest) bright and dark excitons increases now to 95 meV. As such, the weak hybridization between 6T frontier orbitals can enhance the bright-to-dark excitations splitting. Further the dimer bright $S_1$ is redshifted by 63~meV as compared to the monomer $S_1$ energy. 

It is difficult to conduct a fully \textit{ab initio} calculation in the case of a long 1D chain in interaction with the BN tube. Following simple models of dipole-dipole interactions along 1D arrangements, \cite{Hestand2018} the excitonic bandwidth in infinite 1D chains should be twice that of the dimer splitting between the bright and dark combinations, namely about 150 meV in the present case. As such, the lowest eigenvalue should be  located $\sim$75 meV below the monomer S$_1$ energy. Such a value may become somehow larger in the case of non-negligible hybridization between the MOs of neighboring monomers as shown above. This remains  smaller than the $\sim$300 meV redshift observed between a 6T in solution (toluene) and 1D chain of 6T encapsulated in a BN tube  \cite{Juergensen2025}.  Part of the observed 300~meV redshift may however be associated with the evolution of the optical spectrum upon incorporation of the 6T from the solvent to the inner tube.

To substantiate this latter hypothesis, we combine our BSE calculations with a polarizable continuous  model (PCM) of the solvent, adopting the methodology described in Refs.~\citenum{Duchemin2016,Duchemin2018}.  With the dielectric constant and refractive index of toluene, we find that the 6T $S_1$ energy in solution amounts to 2.49 eV, namely 60 meV above the 6T@BN77($\infty$) value accounting for hybridization (Table~\ref{tab:table1}), and 180 meV below the 2.67 eV gas phase value. A qualitatively similar 120 meV redshift from the gas phase to the toluene solution is obtained at the TD-B3LYP/PCM level. As such, part of the observed 300 meV redshift may be associated with the incorporation process from the solvent to the inner-tube location, and not entirely to the 6T-6T interactions in the excited state.  The intercalation of electronically inert (large photoemission and optical gaps) spacers between 6T as a mean to control their distance inside tubes, as elegantly introduced in Ref.~\citenum{Marceau2026}, may provide a refined way to quantify collective effects in the excited state of in-tube  head-to-tail molecular chains. In this latter study, the encapsulation of 6T in the isolated limit (surrounded by spacers)  was shown to induce already a  $\sim$100 meV redshift in the fluorescence emission as compared to 6T in a DMF solvent. 

\section{ Conclusion }

We confirmed the experimental observation that oligothiophene molecules (nT) intercalate easily  inside BN nanotube. The binding energy is optimal  for $\sim$10 $\AA$ diameter tubes,   decaying smoothly for larger tubes, validating the observation that nTs intercalate in BN tubes with a peak distribution in the 10-20~$\AA$ diameter range for the inner-tube  \cite{Badon2023}. In the absence of solvent, we further find that the intercalation in open-ended hydrogen-passivated BN tubes is barrier-less. Further, nTs can slide very easily along the tube, with a corrugation potential much smaller than room-temperature. This is consistent with the observation that nTs may form 1D head-to-tail chains when tubes are too small to accommodate side-by-side nTs  \cite{Badon2023}, as already well characterized in the case of carbon nanotubes  \cite{Gaufres2016}.

Many-body $GW$ and Bethe-Salpeter calculations indicate a rather large impact of the screening by the BN tube electrons on the photoemission and optical gap associated with intercalated nTs. Screening, or polarization, effects can close the photoemission gap by $\sim$750 meV in the case of the 6T@BN(7,7) tube, a value that decreases with tube diameter increase. As expected, the impact of screening on the neutral optical excitations is reduced, still amounting to a $\sim$140 meV redshift for the onset of absorption by the 6T@BN(7,7) system. 

Structural relaxation after intercalation cannot be neglected in the case of the nTs associated with very low energy twisting modes. Upon intercalation, nTs become more planar due to the interaction with the tube walls, inducing an increased conjugation, namely a decreased gap. Such a planarization can result in  a $\sim$100 meV redshift of the lowest singlet absorption energy that adds to screening effects. Hybridization with the tube electronic orbitals is found to be the least sizable effect, with again a redshift of 30-50 meV in the case of the 3T and 6T inside a BN(7,7) tube. Overall, structural relaxation, hybridization and screening can induce a non-negligible 240 meV redshift as compared to the gas phase onset of absorption for the 6T@BN(7,7) system. Such a redshift can be expected to be an upper-bound for incorporation in larger tubes. 

The numbers discussed above are related to intercalated but isolated nTs. We then finally explored the impact of nT-nT interaction in the excited state considering an head-to-tail 6T dimer. Even in the limit of vanishing overlap of the frontier molecular orbitals localized on each monomer, excitonic interactions lead to a 75~meV splitting between a bright and a dark excitonic states, the bright excitation being the lowest in energy. This sizable exciton-exciton interaction is expected to be roughly doubled in the limit of a 1D chain, even though this effect may be reduced by the BN tube screening. The excited state properties of 1D chains inside realistic tubes, accounting for structural relaxation, hybridization, screening, and likely disorder, stands as a significant challenge to theory.


\begin{acknowledgments}
X.B. acknowledges La\"{e}titia Marty for suggesting this study and providing initial bibliography, together with Marivi Fern\'{a}ndez-Serra and Etienne Gaufr\`{e}s for stimulating discussions.
Computational resources were generously provided by  the national HPC facilities under contract GENCI-TGCC A0110910016.
\end{acknowledgments}

\section*{Data Availability Statement}


The data that support the findings of this study are available within the article  and its Supplemental Material.

\appendix*

\section{ Simplified dielectric model for the BN tube susceptibility }

To scan the evolution of the polarization energies as a function of nT length,   tube length and diameter, we set-up a simplified  model susceptibility operator for the BNTs. As shown in a previous study associated with defects in planar \textit{h}-BN  
\cite{Amblard2022},  the polarizability of planar \textit{h}-BN flakes was shown to be perfectly linear with the number of B/N atoms, supporting a  model  where the \textit{h}-BN medium reacts to an external perturbation by local effective atomic induced dipoles. This is consistent with the fact that in insulators, the non-local susceptibility $\chi({\bf r},{\bf r}'; \omega)$ is short-range, namely exponentially decaying as a function of the $| {\bf r}-{\bf r}'|$ distance. We can describe thus the \textit{interacting} susceptibility over a set of auxiliary \textit{p}-orbitals on each nanotube B/N atom, namely an auxiliary basis $\lbrace \; P_{I\mu}({\bf r})=P_{\mu}({\bf r} - R_{I}) \; \rbrace$ where $\mu=x,y,z$ and $R_{I}$ a tube B/N atom. As such:
$$
 \chi({\bf r},{\bf r}') = \sum_{I}^{tube} \sum_{\mu\nu}^{xyz} P_{\mu}({\bf r} - R_{I}^{tube}) \cdot \chi_{\mu\nu}^{at} \cdot P_{\nu}({\bf r} - R_{I}^{tube})
$$
where we restrict our model to the static response limit. In the auxiliary basis, the susceptibility is thus assumed to be diagonal by blocks with the atomic effective susceptibility $\chi_{\mu\nu}^{at}$ 3$\times$3 tensor identical on each atom (we do not attempt to distinguish B and N atomic susceptibilities). In practice,  we take the $\lbrace P_{I\mu} \rbrace$ auxiliary basis set to be very localized Gaussian (l=1) orbitals for simplicity. The polarizability $\alpha_{\mu\nu}$ can then be written:
$$
\alpha_{\mu\nu} = - \sum_{I} \langle x_{\mu} | P_{I\mu} \rbrace \cdot \chi_{\mu\nu}^{at} \cdot \langle P_{I\nu} | x_{\nu} \rangle 
$$
where e.g. $\langle x_{\mu} | P_{I\mu} \rbrace$ is the dipole moment along the $\mu$-direction associated with the $P_{I\mu}$ auxiliary \textit{p}-orbital. The polarizability $\alpha_{\mu\nu}$ is just proportional to the number of B/N atoms, using identical  auxiliary $P_{I\mu}$ orbitals on each atoms.
Taking the RPA \textit{ab initio} data for the polarizability of \textit{h}-BN flakes from Ref.~\citenum{Amblard2022}, one can extract the values of the atomic $\chi_{\mu\nu}^{at} $ tensor. This local atomic polarizability tensor is diagonal for a planar \textit{h}-BN flake, but with the out-of-plane component about 70$\%$ smaller than the in-plane ones. This model atomic susceptibility can be used for nanotubes by a local rotation so that the perpendicular (out-of-plane) direction matches the local normal vector in cylindrical coordinates, yielding an atom dependent local polarizability tensor that we will be writing ${\tilde \chi}_{I\mu\nu}^{at} = U_I^{\dagger} \chi_{\mu\nu}^{at} U_I$, where $U_I$ is the rotation unitary operator bringing the cartesian coordinates into the local cylindrical coordinates. We insist on the fact that this model is not adjusted to reproduce the polarization energy of  nT@BN systems for specific BN tubes with a specific diameter. With such a susceptibility operator, the polarization energy  for a unit source charge localized somewhere in ${\bf R}_s$ in the tube reads
$$
P = -\frac{1}{2} \sum_{I} \sum_{\mu\nu} V_{p\mu}^s( {\bf R}_s - {\bf R}_I) \cdot {\tilde \chi}_{I\mu\nu}^{at} \cdot
V_{p\nu}^s( {\bf R}_s - {\bf R}_I) 
$$
The $V_{p\mu}^s( {\bf R}_s - {\bf R}_I )$ represents e.g. the Coulomb interaction energy between a  source \textit{s}-charge (normalized) located in ${\bf R}_s$ and an (induced) p-dipole oriented along the $\mu$-direction on the B/N atom I. As for induced dipoles, point charges are described by very localized \textit{s}-like (l=0) Gaussian orbitals. With Gaussian charges and dipoles, Coulomb integrals are analytic. In this expression for the polarization energy $P$, one Coulomb integral represents the potential originating from the source charge in ${\bf R}_s$ on the (I) atom of the BN tube. This potential induces on (I) a dipole through the ${\tilde \chi}_{I\mu\nu}^{at}$ susceptibility. The second Coulomb integral represents the reaction field energy of this induced dipole onto the source charge. The (1/2) factor indicates an adiabatic process where the source charge is slowly built (or brought from infinity in the photoemission process). As a last ingredient, the Mulliken charges associated with the HOMO and LUMO levels of the nT are calculated \textit{ab initio} with the {\sc{Orca}} package.  They will be used as the source charges.   Such an approach is a simplified version of generic microelectrostatic models developed  for polarizable atomistic environments (for a review see e.g. Ref.~\citenum{DAvino2016}). While microelectrostatic models introduce effective atomic polarizabilities, the present model takes the susceptibility as the central operator. Further, by fitting directly the \textit{interacting} susceptibility, not the independent-electron one, there is no need to inverse the Dyson equation associated with the self-consistent interaction between induced dipoles. 
\newpage

\vskip 2cm

\begin{thebibliography}{89}
\bibitem [{h0e()}]{h0energy}%
  \BibitemOpen
  \href@noop {} {}\bibinfo {note} {The $\langle H_0 \rangle$ energy is larger than the HOMO-LUMO ev$GW$ gap since the S1 eigenstate shows some weight on transitions at higher energy than the HOMO-LUMO one.}\BibitemShut {Stop}%
\bibitem [{sca()}]{scaling}%
  \BibitemOpen
  \href@noop {} {}\bibinfo {note} {In a nutshell, a source charge inside the tube generates an electric field with a magnitude that scales as $1/R^2$ on B/N tube atoms located at a distance $R$ away. In response, these atoms develop an induced dipole of magnitude $p\simeq \alpha^{at}/R^2$, with $\alpha^{at}$ some effective atomic polarizability. These induced dipoles interact with (stabilize) the added charge with a magnitude $p/R^2 \simeq 1/R^4$.}\BibitemShut {Stop}%
\end{thebibliography}

%


\end{document}